\documentclass[preprint,showpacs,preprintnumbers,amsmath,amssymb]{revtex4}

\usepackage{graphicx}
\usepackage{dcolumn}
\usepackage{bm}

\begin{document}


\title{2D Superconductivity: Classification of Universality Classes \\ by Infinite Symmetry}

\author{Carlo A. Trugenberger}
\email{ca.trugenberger@InfoCodex.com}
\affiliation{%
InfoCodex S.A., av. Louis-Casai 18, CH-1209 Geneva, Switzerland\\
Theory Division, CERN, CH-1211 Geneva 23, Switzerland 
}%


\date{\today}

\begin{abstract}
I consider superconducting condensates which become incompressible 
in the infinite gap limit. Classical 2D incompressible fluids possess the dynamical symmetry of area-preserving diffeomorphisms. I show that the corresponding infinite dynamical symmetry of 2D superconducting fluids is the coset 
${{ W_{1+\infty}  \otimes \overline W_{1+\infty}} \over U(1)_{\rm diagonal}}$, with $W_{1+\infty}$ the chiral algebra of quantum area-preserving diffeomorphisms and I derive its minimal models. These define a discrete set of 2D superconductivity universality classes which fall into two main categories: conventional superconductors with their vortex excitations and unconventional superconductors. These 
are characterized by a broken $U(1)_{\rm vector} \otimes U(1)_{\rm axial}$ symmetry and are labeled by an integer level $m$.
They possess neutral spinon excitations of fractional spin and 
statistics $S = {\theta \over 2\pi} = {{m-1} \over 2m}$ which carry also an $SU(m)$ isospin quantum number; 
this hidden $SU(m)$ symmetry implies that these anyon excitations are non-Abelian. The simplest unconventional superconductor is realized for $m=2$: in this case the spinon excitations are semions (half-fermions).
My results show that spin-charge separation in 2D superconductivity is a universal consequence of the infinite symmetry of the ground state. This infinite symmetry and its superselection rules realize a quantum protectorate in which the neutral spinons can survive even as soft modes on a rigid, spinless charge condensate.   
\end{abstract}

\maketitle

Superconductors are materials in which the electromagnetic U(1) gauge invariance is spontaneously broken. As has been stressed by Weinberg \cite{weinberg}, this is a sufficient condition for superconductivity, irrespective of the model details. As a consequence of the Anderson-Higgs mechanism, density fluctuations in such materials are normally suppressed by a gap. There
are, however, various mechanisms that can lead nonetheless to gapless excitations, notably interactions with magnetic impurities \cite{abrikosov} and nodes in the gap due to p- or d-wave pairing in high-$T_c$ materials \cite{anderson1}. 
These can lead to observable effects in various thermodynamic quantities at low temperatures \cite{leesimon}. 

In this paper I will consider superconductors for which density fluctuations are gapped. The only possible low-lying excitations in this class can involve thus exclusively other degrees of freedom, like e.g. the spin. I will show that, in 2 dimensions (2D), such superconductors fall into universality classes classified by the representations of an infinite-dimensional symmetry algebra. I will argue that, due to the phenomenon of spin-charge separation, these results apply also to the high-$T_c$ cuprates and that the superselection rules of the infinite-dimensional symmetry realize Anderson's quantum protectorate scenario \cite{anderson2}. 

In order to describe the quantum order \cite{wen1} of the superconducting 
ground states, i.e. the universal properties embedded in the quantum entanglement of their wave functions, one can take the infinite-gap limit, in which all
non-universal details are removed. In this limit, the rigidity of the ground-state wave function is also infinite and the system becomes incompressible. 
Since crystals typically do not have the ability to carry currents with no resistive loss, especially in the presence of impurities, I shall concentrate on liquid ground states. 

The classical {\it dynamical symmetry} of incompressible fluids is that of volume-preserving diffeomorphisms. 
These are the transformations that span the whole configuration space when applied to an initial, reference configuration. At the quantum level, dynamical symmetries generate the whole Hilbert space of a theory from a reference quantum state, called the ground state, or vacuum. Therefore, they are also often called spectrum-generating symmetries.

Quantum states are typically organized in highest-weight representations of the symmetry group. If a dynamical symmetry is present, it is possible to organize various irreducible highest-weight representations into consistent sets which are closed under the fusion rules for making composite representations (bootstrap). Such a consistent grouping of representations determines a ground-state with a self-contained set of excitations having well-defined quantum numbers: this data define a {\it universality class} of quantum systems with the given dynamical symmetry. 

Dynamical symmetries are thus ideal for classifying universality classes, the paramount example being conformal field theories \cite{ginsparg}
as universality classes of two-dimensional (2D) critical behaviour \cite{cardy}. This example shows that the classification power of dynamical symmetries is particularly powerful when the symmetry group is infinite-dimensional. In this case, the ground state and the excitations are determined by an infinite number of highest-weight conditions, which lead to significant restrictions on the possible consistent theories. In addition, the quantization of infinite-dimensional symmetry algebras involves subtle modifications leading to the appearance of new quantum numbers, like e.g. the Virasoro central charge in the 2D conformal algebra \cite{ginsparg}. As a consequence the excitations that can arise from this infinite number of degrees of freedom can show quite unexpected emergent behaviour. 

The algebra of volume-preserving diffeomorphisms is infinite-dimensional. One can thus expect a restricted possible set of universality classes of quantum incompressible fluids, defined as consistent theories with the corresponding quantum dynamical symmetry. This program can be explicitly carried out in 2D, where both the unique quantization $W_{1+\infty}$ of the classical algebra $w_\infty$ of area-preserving diffeomorphisms \cite{shen} and its irreducible, highest-weight representations \cite{kac} are well understood.

The best known examples of incompressible fluids are Laughlin's quantum Hall fluids \cite{laughlin}. These are the chiral quantum states leading to the fractional quantum Hall effect \cite{girvinprange}. They were indeed shown to possess a
$W_{1+\infty}$ dynamical symmetry \cite{ctz1} and the Jain hierarchy \cite{jain} of observed quantum Hall states corresponds exactly to the $W_{1+\infty}$ minimal models \cite{ctz2}. These define chiral incompressible fluids which are characterized, loosely speaking, by a minimal set of states and have thus a particularly robust quantum order. 

The algebra $W_{1+\infty}$ is actually the quantization of only one chiral sector of the classical algebra $w_\infty$ and
is thus the appropriate symmetry algebra of chiral incompressible quantum fluids. In order to determine the correct 
dynamical symmetry of P- and T-invariant superconducting fluids I will therefore start from the direct product of two copies $W_{1+\infty}$ and $\overline W_{1+\infty}$ of opposite chirality. This is the quantum version of the full classical $w_\infty$ algebra. The spontaneously broken $U(1)$ gauge symmetry can then be taken into account as follows. The superconducting charge condensate which breaks the $U(1)$ gauge symmetry has the consequence that charge ceases to be a good quantum number in a superconducting ground state. The quantum algebra $W_{1+\infty}$ contains a $\widehat U(1)$ Kac-Moody current \cite{ginsparg} $V^0_n$, so that the diagonal vector current $V^0_n + \overline V^0_n$ is naturally identified with the electric charge current. This Kac-Moody symmetry has therefore to be divided out from the dynamical symmetry group, leaving the coset
\begin{equation}
W = {{W_{1+\infty} \otimes \overline W_{1+\infty}} \over  \widehat U(1)_{\rm diagonal}} \ .
\label{aa}
\end{equation}
This construction is fully supported
by a recent result \cite{dst} about global superconductivity in planar Josephson junction arrays. This is indeed realized as a coset topological fluid described by the axial combination of two Chern-Simons gauge fields of opposite chiralities \cite{wen2}, the residual symmetry in the broken phase being
$U(1) \otimes \overline U(1)/ U(1)_{\rm diagonal} = U(1)_{\rm axial}$. As I will show below this is the simplest realization
of W. 

Having determined the relevant dynamical symmetry, one can classify possible 2D superconducting fluids as consistent W-theories, characterized by the quantum numbers and degeneracies of their excitations. 
Actually, I propose to classify 2D superconductivity universality classes as {\it W minimal models}. These are a particular subset of W-theories with less states than the generic theories with the same symmetry. 
It is rather natural that the theories with a minimal set of excitations possess a particularly robust quantum order, since
many "channels" of perturbation are missing. Consider, e.g., perturbing the ground state by raising the temperature from zero to a finite value. Since the energy levels can be expected to be generically continuous functions of the model parameters, 
the energy levels of minimal models will be only marginally different from those of the neighbouring generic models. However, they correspond to special values of the continuous model parameters for which many of the excited states of the neighbouring generic models are simply missing. Therefore, minimal models have a smaller partition function than their neighbouring generic models and, correspondingly, the probability of finding the system still in the ground state at finite temperature, which has the partition function in the denominator, has a local maximum at the minimal models. Minimal models
correspond thus to special parameters for which the quantum order of the ground state becomes particularly robust. 
This long-distance stability principle leads to a logically self-contained theory of 2D superconductivity with far-reaching consequences, as I now show.  

Area-preserving diffeomorphisms are canonical transformations of a two-dimensional phase space. By endowing the plane with coordinates $z$ and $\bar z$ with a Poisson bracket
\begin{equation}
\{ f,g \} = i \ ( \partial f \bar \partial g - \bar \partial f \partial g )\ ,
\label{ab}
\end{equation}
one can describe area-preserving diffeomorphisms $\delta z = \{ {\cal L}, z \}$ and $\delta \bar z =\{ {\cal L}, \bar z \}$ in terms of generating functions ${\cal L}(z, \bar z)$. The basis of generators ${\cal L}_{n,m} = z^n \bar z^m$
satisfy the classical $w_{\infty}$ algebra \cite{shen}
\begin{equation}
\left\{ {\cal L}_{n,m}, {\cal L}_{k,l} \right\} = -i \ (mk-nl) \ {\cal L}_{n+k-1,m+l-1} \ .
\label{ac}
\end{equation}
Note that the operators ${\cal L}_{nm}$ with $n\ge 0$ and $m\ge 0$ form two closed sub-algebras related by complex conjugation. These are the two chiral sectors of the classical $w_{\infty}$ algebra. Generators with both $n$ and $m$ negative 
are descendants that can be obtained as products of primary generators in the two fundamental chiral sectors. 

The quantum version of this infinite-dimensional algebra is obtained by the usual substitution of Poisson brackets with quantum
commutators: $i\{ ,\} \rightarrow [ , ]$. In order to make contact with the standard notation I shall denote the quantum
version of ${\cal L}_{i-n,i}$ by $V^i_n$. By restricting to positive values of $i$ I discuss first one chiral sector of the quantum algebra, 
\begin{equation}
{[ V^i_n, V^j_m]} = (jn-im) \ V^{i+j-1}_{n+m} +q(i,j,n,m)\ V^{i+j-3}_{n+m}
+\cdots +\delta^{ij}\delta_{n+m,0}\ c\ d(i,n) \ ,
\label{ad}
\end{equation}
where the structure constants $q$ and $d$ are polynomials of their arguments,
and the dots denote a finite number of
similar terms involving the operators $V^{i+j-1-2k}_{n+m}$. The first term on the r.h.s. of (\ref{ad}) is the classical term
(\ref{ac}). The remaining terms are quantum operator corrections, with the exception of the last c-number term, which represents a quantum anomaly with central charge $c$. All the quantum operator corrections are uniquely determined by the closure of the algebra; only the central charge $c$ is a free parameter. The quantization of the full classical algebra
$w_{\infty}$, involving generators in the two chiral sectors and all their products, is then obtained as the direct product
of two copies $W_{1+\infty}$ and $\overline W_{1+\infty}$ of opposite chirality.

The generators $V^i_n$ are characterized by an integer conformal (scaling) dimension $h=i+1 \ge 1$ and an angular momentum mode index $n$,
$-\infty < n < +\infty $. 
The operators $V^0_n$ satisfy the Abelian current algebra (Kac-Moody algebra)
$\widehat U(1)$ \cite{ginsparg} while the operators $V^1_n$ are the generators of conformal transformations, satisfying the Virasoro algebra \cite{ginsparg},
\begin{eqnarray}
{[V^0_n,V^0_m]}    && =  n \ c\ \delta_{n+m,0} \ ,
\nonumber \\
{[ V^1_n, V^0_m }] && =  -m\ V^0_{n+m} \ ,
\nonumber \\
{[ V^1_n, V^1_m ]} && =  (n-m)V^1_{n+m} +{c\over 12}n(n^2-1)
\delta_{n+m,0}\ .
\label{ae}
\end{eqnarray}
$V^0_n$ and $V^1_n$ are identified as the charge and angular-momentum modes
in the chiral sector under consideration. 

A $W_ {1+\infty}$-theory is defined as a Hilbert space constructed as a set of
irreducible, unitary, highest-weight representations of $W_ {1+\infty}$,
which is closed under the fusion rules for making composite states.
The irreducible, unitary, quasi-finite, highest-weight
representations of $W_ {1+\infty}$ have been completely classified in \cite{kac}. 
They exist only if the central charge is a positive
integer, $c=m$, $m\in {\bf Z}_+$.
They are characterized by an $m$-dimensional weight vector 
$\vec{r}$ with real elements and are built on top of
a highest weight state $|\vec{r}\rangle_W$ which satisfies
\begin{equation}
V^i_n|\vec{r}\rangle_W = 0 \ ,\qquad \forall\ n>0\ , \ i \ge 0 \ ,
\label{af}
\end{equation}
and is an eigenstate of the $V^i_0$,
\begin{equation}
V^i_0|\vec{r}\rangle_W =\sum_{n=1}^m \ m^i (r_n)\ |\vec{r}\rangle_W \ ,
\label{ag}
\end{equation}
where $m^i(r)$ are $i$-th order polynomials of a weight component. In particular, the charge $V^0_0$ and scaling dimension $V^1_0$
are given by
\begin{eqnarray}
q = \sum_{n=1}^m \ m^0 (r_n) &&= r_1 + \cdots + r_m\ ,
\nonumber\\
h = \sum_{n=1}^m \ m^1 (r_n) &&= {1\over 2}\left[
  \left(r_1 \right)^2 +\cdots + \left(r_m \right)^2 \right]\ .
\label{ah}
\end{eqnarray}
Note that scaling dimension and angular momentum coincide in a chiral theory. 

In this algebraic construction, the incompressible quantum fluid ground state is the special
highest-weight state $|\Omega\rangle_W$ satisfying
\begin{eqnarray}
V^i_n \vert\Omega\rangle_W &&=0\ , \quad \forall\ n >0\ ,\ i\ge 0\ ,
\nonumber \\
V^i_0 \vert\Omega\rangle_W &&=0\ ,\ i\ge 0\ .
\label{ai}
\end{eqnarray}
Particle-hole excitations above the ground state are obtained by applying
generators with negative mode index to $\vert\Omega\rangle_W$. Due to incompressibility, these low-lying
excitations are small deformations of the sample boundary and are thus pure edge excitations. 
Bulk excitations are identified with the other highest-weight representations in the $W_ {1+\infty}$-theory
and are characterized by an infinite set of quantum numbers (\ref{ag}). Each of these excitations has its own tower
of low-lying edge excitations. Since the representations that I consider are quasi-finite, the number $d(n)$ of independent edge excitations at total angular momentum level $n$, encoded in the character of the representation, is finite.

There are two types of irreducible, unitary, quasi-finite, highest-weight
representations of $W_ {1+\infty}$ \cite{kac}. For {\it generic} representations the weight vector
$\vec{r}$ is such that $(r_i-r_j)\not\in {\bf Z},\ \forall\ i\neq j$. These representations 
are equivalent to the corresponding $\widehat U(1)^{\otimes m}$ representations with the same weight. 
{\it Degenerate} representations, instead, have 
$(r_i-r_j) \in {\bf Z}$ for some $i\neq j$.
The weight components $\{r_i\}$ of the degenerate representations
can be grouped and ordered in congruence classes modulo ${\bf Z}$
\cite{kac}. A representation with two classes is the tensor product of two one-class
representations. Therefore, the one-class degenerate
representations are actually the basic building blocks for all degenerate representations. 
These one-class representations have the weight vectors
\begin{equation}
\vec{r} = \{r_1,\dots,r_m\} = \{s+n_1,\dots,s+n_m\}\ ,\quad s\in {\bf R}\, \quad
n_1\ge\cdots \ge n_m \in {\bf Z}\ .
\label{aj}
\end{equation}

The $c=m$ one-class degenerate $W_{1+\infty}$ representations are in one-to-one relation with the
representations of $U(1) \otimes SU(m)$. Indeed, by an orthogonal transformation it is possible to
introduce \cite{ctz2} a new basis $\vec{q}=\{q,{\bf\Lambda}\}$ for the $W_{1+\infty}$ weights:
\begin{eqnarray}
q &&= {1\over\sqrt{m}} \ \left( r_1+r_2+\cdots +r_m \right) \ ,
\nonumber\\
\Lambda_a &&= \sum_{i=1}^m\ u^{(i)}_a \ r_i\ ,\qquad a=1,\dots, m-1\ ,
\label{al}
\end{eqnarray}
where ${\bf u}^{(i)}$ are the weight vectors of the defining $SU(m)$ representation \cite{georgi}.
Each such representation embodies therefore one charged excitation and $(m-1)$ neutral excitations with a hidden $SU(m)$ symmetry. 

In the $\{q,\Lambda\}$ basis the fusion rules for making composite $W_{1+\infty}$ representations take a particularly
simple form:
\begin{eqnarray}
\vec{q} \bullet \vec{p} &&= \vec{p}+\vec{q} \qquad
{\rm mod}\quad \left\{ {0 \choose {\bf\alpha}^{(1)} }\ ,\cdots,
 {0 \choose {\bf\alpha}^{(m-1)} } \right\} \ ,
\nonumber \\
{\bf\alpha}^{(a)} &&= {\bf u}^{(a)} -{\bf u}^{(a+1)}\ ,\qquad a=1,\dots ,m-1 \ .
\label{an}
\end{eqnarray}
This means that charge is additive, while the neutral excitations combine according to the
$SU(m)$ fusion rules, i.e. the $SU(m)$ weights add up modulo an integer combination of the simple
roots ${\bf\alpha}^{(a)}$.

The number $d(n)$ of independent edge excitations at level $n$ is lower for
degenerate representations than for generic ones,
because the former have additional relations among the states,
leading to null vectors which have to be projected out in order to maintain
irreducibility \cite{ginsparg}. 
This is the origin of the reducibility of the $\widehat U(1)^{\otimes m}$
representations with respect to the $W_{1+\infty}$ algebra.
On the other hand, the one-class degenerate $W_{1+\infty}$ representations are
one-to-one equivalent to those of the
$\widehat U(1) \otimes {\cal W}_m $ minimal models, where
${\cal W}_m$ is the Fateev-Lykyanov-Zamolodchikov algebra \cite{fz} in the limit $c_{{\cal W}_m} \to m-1$.

This minimality of states should be interpreted as a long-distance stability principle
which ensures particularly robust ground states by way of a minimal set of 
perturbation channels, as explained above. 
Encouraged by the success of this principle
in predicting exactly the observed hierarchy of quantum Hall states \cite{ctz2} I propose 
to apply it also to the classification of 2D superconducting fluids. Henceforth I shall thus concentrate only on 
the $W_{1+\infty}$ {\it minimal models}, defined as $W_{1+\infty}$-theories made only of one-class degenerate
representations. 

These minimal models were constructed in \cite{ctz2}. They are made by weights belonging to lattices $L$ which are closed
under the fusion rules (\ref{an}). In the original $\vec{r}$ basis,
these lattices $\vec{r}=\sum_{i=1}^m n_i\ \vec{v}_i\ $ are generated by the basis vectors
\begin{equation}
\left( \vec{v}_i\right)_j =\delta_{ij} + rC_{ij} \ ,
\label{ap}
\end{equation}
where $r \in {\rm R}$, $C_{ij} = 1$, $\forall i,j = 1\dots m$ and the integer excitation labels $n_i$ must obey
the constraints $n_1 \ge n_2 \ge \dots \ge n_m$ in order to avoid double counting. The real number $r$ derives from the
free parameter $s$ of the one-class degenerate representations in (\ref{aj}).  

Using the basis (\ref{al})and the fact that $\sum_{i=1}^m {\bf u}^{(i)} = 0$ it is easy to recognize that the 
excitations having no neutral component are of the form $n_i = n$, $\forall i = 1\dots m$. This defines the $\widehat U(1)$
axis of the lattice. The charge unit is thus given
by the lattice vector $\sum_{i=1}^m \vec{v}_i$ and the charge of a generic excitation is the linear projection of its lattice vector on this unit charge vector. The charge and scaling dimension of the excitation with label $\{ {\bf n} \}$ are thus given by
\begin{eqnarray}
q &&= {\bf t}^T \cdot M \cdot {\bf n} \ , 
\nonumber\\
h  &&= {1\over 2} \ {\bf n}^T \cdot M\cdot {\bf n} \ ,
\label{aq}
\end{eqnarray}
where ${\bf t}=(1,\dots ,1)$ and the matrix $M$ is the metric of the lattice $L$:
\begin{equation}
M_{ij} = \vec{v}_i \cdot \vec{v}_j =\delta_{ij} +\lambda \ C_{ij} \ ,
\quad \lambda =mr^2 + 2r \ .
\label{ar}
\end{equation}
As a consequence, the charge unit of the theory is determined by the parameter $r$ as $q_{\rm unit}= m \left(
1+mr \right)^2$. Note that the spectrum of the theory contains also fractionally charged excitations. 

Having completed the construction of the chiral minimal models I turn now to the conjugate sector of opposite chirality. 
This is spanned by the complex conjugate generators $\bar z^{i-n}z^i$ with $i \ge 0$. 
Their algebra can be explicitly computed using the same quantum commutator $[z, \bar z]
=-1$ that lead to (\ref{ad}). It differs from $W_{1+\infty}$ by an overall sign. 
The generators $\overline V^i_n = (-1)^i\ \bar z^{i-n}z^i$ satisfy, thus, exactly the 
same algebra (\ref{ad}) as the original operators $V^i_n$: I will call this algebra of opposite chirality
$\overline W_{1+\infty}$.  

A weight vector $(r_1, \dots , r_n)$ of $W_{1+\infty}$ is also a weight vector of $\overline W_{1+\infty}$ and the
weights $\bar m$, $\bar q$ and $\bar h$ are given by the same polynomial expressions (\ref{ag}) and (\ref{ah}). In this
sector of opposite chirality, however, the angular momentum coincides with the negative $-\bar h$ of the scaling dimension.
A minimal model of $W_{1+\infty} \otimes \overline W_{1+\infty}$ at level $c=m$ is thus spanned by a lattice $L \otimes L$ with $L$ defined in
(\ref{ap}). Its excitations have integer labels $\{ {\bf n}, {\bf \bar n} \}$ and charge, vorticity, spin and scaling quantum numbers
\begin{eqnarray}
Q &&= q + \bar q \ , \quad \Phi = q-\bar q \ ,
\nonumber \\
S &&= h - \bar h \ , \quad H = h + \bar h \ .
\label{au}
\end{eqnarray}

This is not yet the correct dynamical symmetry of a superconducting fluid however. A superconducting ground state, in fact, breaks the 
$U(1)$ gauge symmetry, so that charge ceases to be a good quantum number. In my formalism, this can be incorporated by dividing out from the 
dynamical symmetry algebra the diagonal $\widehat U(1)$ Kac-Moody algebra identified with the electric charge current,
leading to the coset algebra 
\begin{equation}
W = {{W_{1+\infty} \otimes \overline W_{1+\infty}} \over \hat U(1)_{\rm diagonal}} \ .
\label{av}
\end{equation}
This can be accomplished by restricting to lattices $L$ for which the total charge vanishes identically, $Q=0$. 
Then there are no charged excitations in the spectrum and the generators $V^0_{-n} +\overline V^0_{-n}$ can be consistently eliminated from the construction of edge excitations. 

Using eqs. (\ref{au}) and (\ref{aq}) it is easy to rewrite the condition $Q=0$ as $Q=\left( 1+mr \right) ^2 \sum_i \left( n_i + \bar n_i \right) = 0$. There are two solutions to this equation. The first is a generic solution and consists of restricting to excitations with quantum numbers
$\{ {\bf n}, {\bf \bar n} = {\bf n}^R\}$, where the reflected excitation $R$ is defined by $n_1^R = -n_m, \dots n_m^R =-n_1$, so that both constraints
$Q=0$ and $\bar n_1 \ge \bar n_2 \ge \dots \ge \bar n_m$ are satisfied. 
In this case, the full diagonal $\left( W_{1+\infty} \right)_{\rm diagonal}$ is broken
by the superconducting ground state and the dynamical symmetry reduces to
\begin{equation}
W = \left( W_{1+\infty} \right)_{\rm axial} \ .
\label{aw}
\end{equation}
The excitation spectrum reduces to bosonic vortices with fluxes $\Phi = k \Phi_0$ which are integer multiples of the flux quantum $\Phi_0 = 2(1+mr)^2$ and have an internal structure with momenta determined by the eigenvalues of the 
higher $\left( W_{1+\infty} \right)_{\rm axial}$ generators. 
I will call solutions with this
highly reduced dynamical symmetry "conventional supercondutors". Conventional superconductors are the only solution at
level $c=1$, where the full dynamical symmetry reduces essentially to $\widehat U(1)_{\rm axial}$: as was recently shown in
\cite{dst} this universality class is realized in planar Josephson junction arrays. 

The most interesting situation occurs, however, when the parameter $r$ assumes the value $r=-1/m$. In this case the chiral lattice $L$ is degenerate and 
consists entirely of neutral excitations, so that no additional conditions have to be imposed in order to divide out the diagonal subgroup $\widehat U(1)_{\rm diagonal}$. Actually, this has the consequence that also the vorticity $\Phi$ vanishes identically, so that both $\widehat U(1)_{\rm vector}$ and $\widehat U(1)_{\rm axial}$ are broken by the ground state and 
the effective dynamical symmetry becomes
\begin{equation}
W = W_m \otimes \overline W_m \ ,
\label{ay}
\end{equation}
with $W_m$ the Fateev, Lykyanov, Zamolodchikov algebra \cite{fz} in the limit $c_{{\cal W}_m} \to m-1$. 
I will call these second series of minimal models "unconventional superconductors". 

The spectrum of unconventional superconductors consists entirely of neutral spinon excitations \cite{fradkin} with spin
\begin{eqnarray}
S &&= h-\bar h \ ,
\nonumber \\
h &&= {1\over 2} \ \sum_i n_i^2 - {1\over m} \left( \sum_i n_i \right) ^2 \ ,
\nonumber \\
\bar h &&= {1\over 2} \ \sum_i \bar n_i^2 - {1\over m} \left( \sum_i \bar n_i \right) ^2 \ .
\label{ax}
\end{eqnarray}
This is the phenomenon of {\it spin-charge separation} \cite{anderson1}: the charge and spin degrees of freedom 
of the original constituents split into two separate entities, charge condenses into a superconducting state and spin resides on neutral quasi-particle excitations about this state. 
Here I have shown that spin-charge separation is a universal feature of 2D superconductivity which follows uniquely from the infinite symmetry of superconducting fluids, independently from the details of the various models. 

A particularly suitable basis for the $(m-1)$ elementary excitations is given by the
vectors $({\bf n}^{(a)},{\bf 0})$ with ${\bf n}^{(a)}$ 
defined by  $n^{(a)}_i = 1$ $\forall i \le a$ and $n^{(a)}_i = 0$ $\forall i  > a$. The corresponding weight vectors
in the basis (\ref{al}) become then the $SU(m)$ fundamental weights,
\begin{equation}
{\bf \Lambda} = {\bf \Lambda}^{(a)}= \sum_{i=1}^a {\bf u}^{(i)} \ .
\label{az}
\end{equation}
Each elementary excitation is thus associated with the heighest weight of an $SU(m)$ fundamental representation and has spin
\begin{equation}
S^{(a)} = {a\over 2} \left( 1- {a\over m} \right) \ .
\label{aaa}
\end{equation}
the lowest possible value being 
\begin{equation}
S_{\rm min}={{m-1} \over 2m} \ .
\label{aab}
\end{equation}
The conjugate excitations of opposite chirality have the same structure with spins of opposite sign.

The spinons are thus generically anyons \cite{wilczek} with fractional statistics $\theta/\pi = 2S$ (exp(i$\theta$) is the
particle-exchange factor), the simplest example being the semions (half-fermions) of the lowest level ($c=2$) unconventional superconductor.  
These anyons carry an $SU(m)$ isospin quantum number: their fractional statistics is therefore
non-Abelian. Note, however, that the $SU(m)$ symmetry of these excitations is different from the usual symmetry of, say, the
quark model of strong interactions \cite{georgi}. Indeed, spinons do not come in full $SU(m)$ multiplets; rather, only the
highest-weight states are present. These, however, combine according to the usual $SU(m)$ fusion rules, which explains the
non-Abelian character of their monodromies. 

To conclude it remains to be discussed if some of the unconventional superconductors predicted by the above symmetry
arguments have already been observed. Indeed, given the central role often assigned to spin-charge separation in high-$T_c$
superconductivity \cite{anderson3}, it is tempting to identify the simplest universality class of unconventional
superconductors ($c=2$) with the high-$T_c$ cuprates. Such an identification, however, is not straightforward and requires justifications. Indeed, it is by now well established \cite{anderson1} that in these materials the pairing occurs with
$d$-wave symmetry, leading to nodes in the gap. This has the consequence that the quasi-particle spectrum contains low-energy
excitations, with observable consequences \cite{leesimon}. The presence of these gapless excitations might seem, at first, to invalidate the assumption of incompressibility in the infinte gap limit. In Anderson's quantum protectorate scenario \cite{anderson2} arising from spin-charge separation and subsequent condensation of the charge degrees of freedom, however, the gapless excitations involve exclusively the spin degree of freedom and do not correspond to charge density fluctuations. This is perfectly compatible with the physical picture of dynamical symmetry under quantum area-preserving diffeomorphisms, which requires only the absence of gapless density fluctuations. Not only it is compatible but it is also exactly the picture that is predicted by W symmetry. The superselection rules of the infinite-dimensional dynamical symmetry would then provide a natural "quantum protector" for the ground state and its low-lying excitations. Note also that the fact that the $\widehat U(1)_{\rm axial}$ group coupled to magnetic flux is broken for the unconventional minimal models naturally suggests that these ground states consists of a condensate of charges paired in higher angular momentum states, a further confirmation of the possible relevance of these models for the high-$T_c$ superconductors.

\end{document}